\documentclass[aps,pre,twocolumn,groupedaddress,showpacs]{revtex4}
\usepackage{graphicx}
\usepackage{amssymb,amsfonts,amsmath}

\newcommand{\ii}{i}          
\newcommand{\cc}{\text{c.c.}}
\newcommand{\eps}{\varepsilon}
\newcommand{\bigO}{\ensuremath{{\mathcal{O}}}}

\newcommand{\tave}[1]{\langle #1 \rangle}
\newcommand{\shorttave}[1]{\langle #1 \rangle_{\tau}}
\newcommand{\expa}{\alpha_1}
\newcommand{\expb}{\beta}
\newcommand{\mone}{\alpha}
\newcommand{\mtwo}{\gamma}
\newcommand{\fscal}{G}       
\newcommand{\fmean}{g}
\newcommand{\fmeanL}{g^{(L)}}
\newcommand{\ffluct}{\tilde{f}}
\newcommand{\sfront}{s}
\newcommand{\Xscal}{Y}
\newcommand{\Li}{L_0}
\newcommand{\Lii}{L_1}
\newcommand{\Liii}{L_2}

\begin{document}

\title{Coarsening to Chaos-Stabilized Fronts}

\author{Ka-Fai Poon}
\altaffiliation{Current address: School of Earth and Ocean Sciences,
  University of Victoria, Victoria, BC V8W 3V6, Canada}
\author{Ralf W. Wittenberg}
 \email{ralf@sfu.ca}
 \affiliation{Department of Mathematics, Simon Fraser University,
   Burnaby, BC V5A 1S6, Canada}

\date{\today}

\begin{abstract}  
 
  We investigate a model for pattern formation in the presence of
  Galilean symmetry proposed by Matthews and Cox [Phys.\ Rev.\ E
  \textbf{62}, R1473 (2000)], which has the form of coupled
  generalized Burgers and Ginzburg-Landau-type equations.  With only
  the system size $L$ as a parameter, we find distinct ``small-$L$''
  and ``large-$L$'' regimes exhibiting clear differences in their
  dynamics and scaling behavior.  The long-time statistically
  stationary state contains a single $L$-dependent front, stabilized
  globally by spatiotemporally chaotic dynamics localized away from
  the front.  For sufficiently large domains, the transient dynamics
  include a state consisting of several viscous shock-like structures
  which coarsens gradually, before collapsing to a single front when
  one front absorbs the others.

\end{abstract}


\maketitle

In the exploration of the rich and diverse range of spatiotemporal
dynamics observed in nonlinear, nonequilibrium spatially extended
systems, it has proved particularly fruitful to investigate
comparatively simple model partial differential equations (PDEs) whose
solutions capture the essential features of the phenomena under
investigation.  Thus the Burgers equation has been extensively studied
for the evolution and statistics of shocks; the Ginzburg-Landau (GL)
equation and its generalizations describe the dynamics and stability
of modulations of patterned states, while the Kuramoto-Sivashinsky and
other models display spatiotemporal chaos (STC) \cite{CrHo93}.  In
this paper we discuss a system describing the amplitude evolution for
pattern formation with symmetry, which appears to combine features of
several of these canonical systems and displays a surprising wealth of
behaviors.


We investigate the Matthews-Cox (MC) equations \cite{MaCo00}
\begin{eqnarray}
  \label{eq:MC1}
  A_T & = & A + 4 A_{XX} - \ii f A , \\
  \label{eq:MC2}
  f_T & = & f_{XX} - |A|^2_X 
\end{eqnarray}
on a one-dimensional $L$-periodic domain, where $A$ is complex, $f$ is
real, and $f_X \equiv \partial_X f \equiv \partial f / \partial X$
(similarly for the other derivatives).  
Equations \eqref{eq:MC1}--\eqref{eq:MC2} were initially derived in
the context of the Nikolaevskiy PDE
\begin{equation}
  \label{eq:nik}
  u_t + u u_x = - \partial_x^2 \left[ \eps^2 - \left( 1 + \partial_x^2
    \right)^2 \right] u  .
\end{equation}
This equation, proposed originally to model seismic wave behavior in
viscoelastic media \cite{BeNi93}, and subsequently obtained in other
contexts \cite{Tana04,CoMa07}, appears to be a canonical model for
short-wave pattern formation with reflection and Galilean symmetries.
Unlike in more common pattern-forming contexts described at onset by
the GL equation, the $\bigO(\eps)$ stationary rolls in \eqref{eq:nik}
are all unstable for all $\eps > 0$
\cite{TrVe96,TrTs96,MaCo00,CoMa07}.  Instead, solutions of
\eqref{eq:nik} exhibit spatiotemporal chaos with strong scale
separation \cite{Tana05,WiPo09}, with coupling between the weakly
unstable pattern at wave numbers $k \approx 1$ and the neutrally
stable long-wave mode with $k \approx 0$.  This suggests the Ansatz
$u(x,t) \sim \eps^{\expa} A(X,T) e^{\ii x} + \cc + \eps^{\expb} f(X,T)
+ \dots$ for the envelopes $A$ and $f$ of the pattern and long-wave
modes, respectively, where $X = \eps x$, $T = \eps^2 t$.  Matthews and
Cox \cite{MaCo00} showed that the asymptotically self-consistent
scaling as $\eps \to 0$ is $\expa = 3/2$, $\expb = 2$, and hence
derived \eqref{eq:MC1}--\eqref{eq:MC2} from the Nikolaevskiy PDE as
the leading-order modulation equations.  While the scaling behavior on
the attractor of \eqref{eq:nik} may be insufficiently described by
this Ansatz \cite{WiPo09}, the MC equations deserve study in their own
right as generic amplitude equations for pattern formation with these
symmetries \cite{MaCo00}.  Since \eqref{eq:MC2} preserves the spatial
mean of $f$, by Galilean invariance we may assume $f$ to have mean
zero.


\paragraph{Chaos-stabilized fronts:}

In describing properties of the MC equations
\eqref{eq:MC1}--\eqref{eq:MC2}, we  emphasize the dynamics of the
large-scale mode $f$, since the pattern amplitude $A$ appears to be
driven by $f$.  We note that several aspects of the behavior for
relatively small $L$ have been previously described by Sakaguchi and
Tanaka \cite{SaTa07}.

\begin{figure}[b]
  \begin{center}
      \includegraphics[width=3.2in]{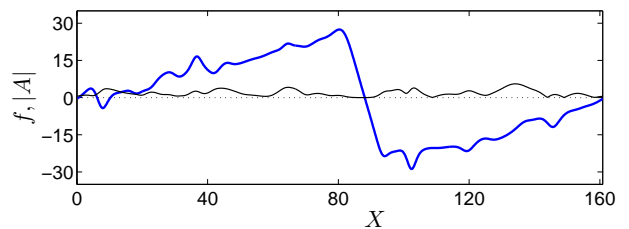}
      \caption{Snapshot at a fixed time $T_1 = 28000$ of $f(X,T_1)$
        (thick blue line) and $|A(X,T_1)|$ (thin black line) for a
        solution of \eqref{eq:MC1}--\eqref{eq:MC2} with $L = 51.2\pi
        \approx 160.8$.\\[-6ex]}
      \label{fig:snapshot}
  \end{center}
\end{figure}

\begin{figure}
  \begin{center}
    \begin{tabular}{cc}
      \includegraphics[width=1.7in]{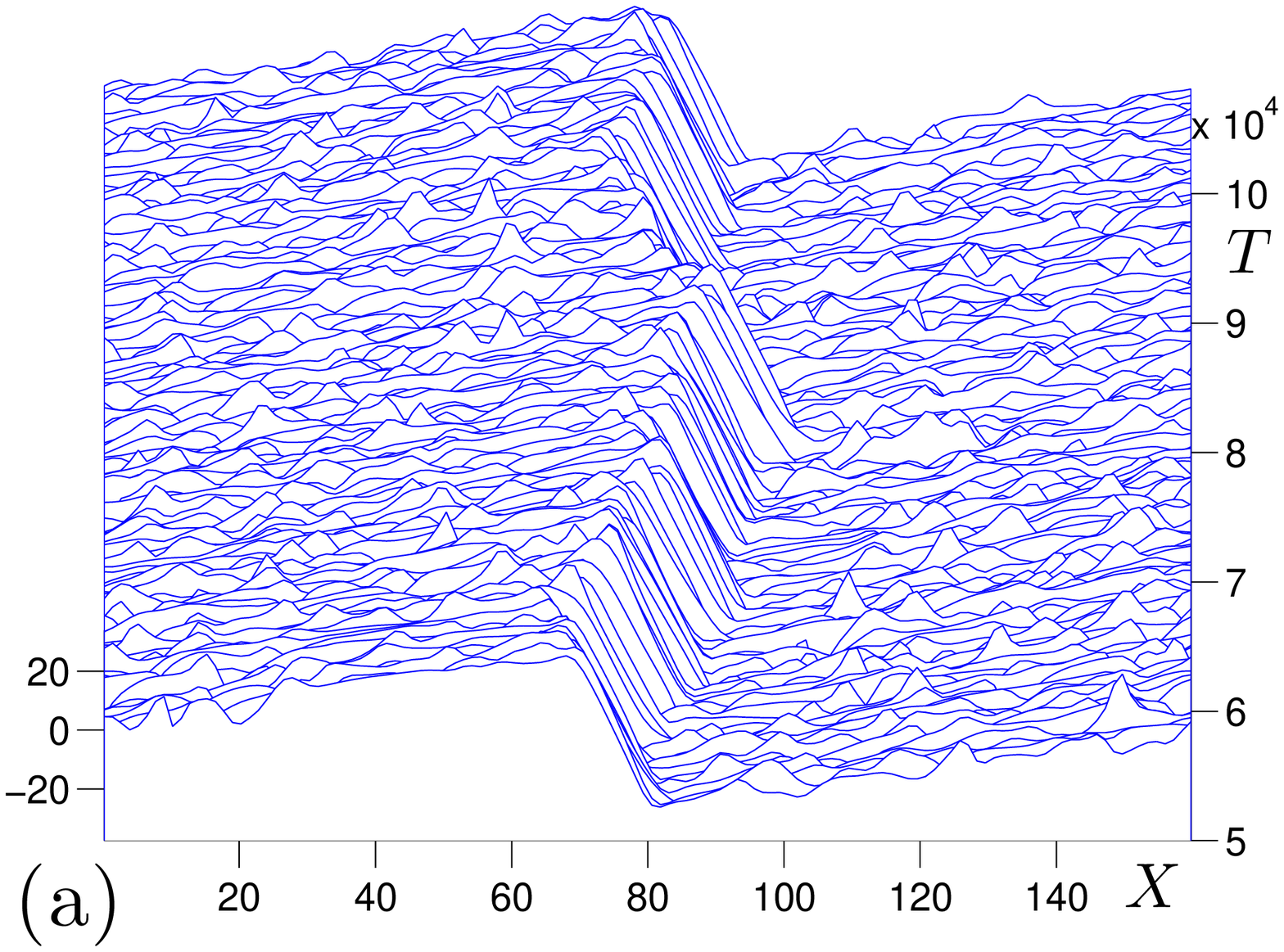}
      & 
      \includegraphics[width=1.7in]{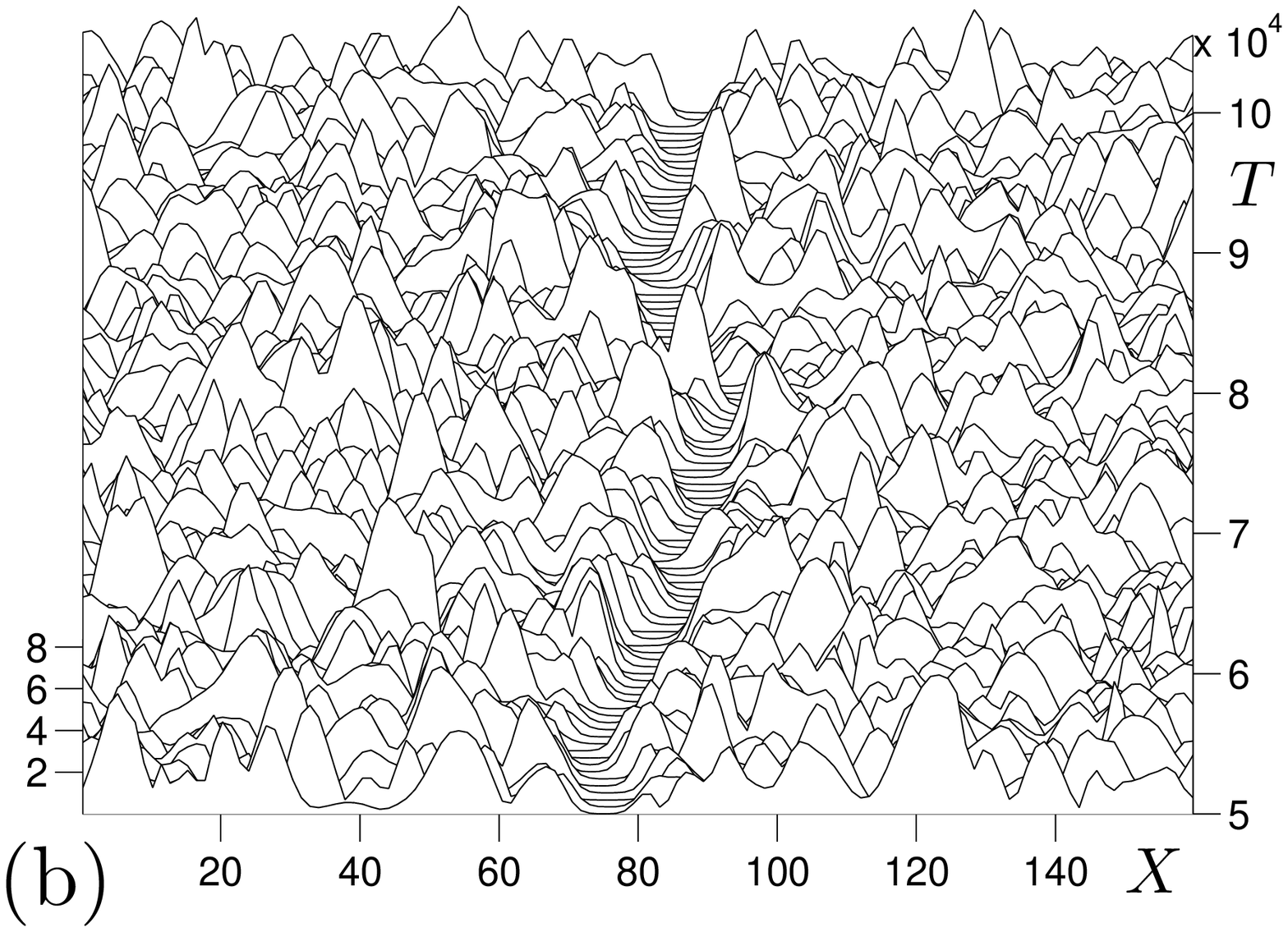}
    \end{tabular}
    \caption{Space-time plots of long-time solutions (a) $f(X,T)$ and
      (b) $|A(X,T)|$ of the MC equations for $L = 51.2\pi$.\\[-6ex]}
    \label{fig:fAxt}
  \end{center}
\end{figure}

The snapshot of a solution for domain size $L = 51.2\pi$
shown in Fig.~\ref{fig:snapshot}
is typical of the statistically stationary behavior for ``small''
domains.  The overall structure of $f$ resembles a perturbed viscous
shock, with $f$ decreasing essentially linearly within the ``front''
region.  Simultaneously, $|A|$ vanishes in the center of the front;
Sakaguchi and Tanaka hence call this an ``amplitude death'' state
\cite{SaTa07}.  The time evolution of a typical solution shown in
Fig.~\ref{fig:fAxt} clearly shows the invariance of the front
structure in $f$ and the suppression of the roll amplitude $A$ within
the front region.
On the remainder of the domain, chaotic fluctuations in $f$ (on
$\bigO(1)$ time scales) are superimposed on the approximately linear
average positive slope, and correlated (in space and time)
with  chaos in $A$.  
This coexistence of an ordered front (amplitude death state) and a
spatially localized spatiotemporally chaotic region is robust on all
domains $L \gtrsim \Li$ large enough to sustain the front \cite{SaTa07}.

The space-time plot indicates that the overall viscous shock profile
in $f$ is nonstationary, but maintains its shape up to small
fluctuations; that is, short-time averages \footnote{Short-time
  averages $\shorttave{f(X,\cdot)}$ are taken over time intervals
  $\tau$ long relative to the $\bigO(1)$ time scales of the chaotic
  fluctuations, but short compared with front translations or
  transient coarsening (we use $\tau \gtrsim 40$).}
$\shorttave{f(X,\cdot)}$ are invariant up to translation.  Denoting
the averaged profile by $\fmean(X) = \fmeanL(X)$, where we center the
front so $\fmean(L/2) = 0$, $\fmean_X(L/2) < 0$, and defining the
front displacement $\sfront(T)$ so that the instantaneous front
position is $X_s(T) = L/2 + \sfront(T)$, we may decompose the
large-scale mode as $f(X,T) = \fmean(X - \sfront(T)) + \ffluct(X -
\sfront(T),T)$, where $\ffluct$ denotes fluctuations about the mean
profile.  The unsteady dynamics in $A$ and $\ffluct$ are then
essentially confined to the region where $\fmean_X \geq 0$ and to the
vicinity of the local extrema of $\fmean$ (see
Fig.~\ref{fig:fmeanL}(c) below).

To help clarify this unusual behavior, we observe that the equation
\eqref{eq:MC2} for the large-scale mode $f$ has the form of a
conservation law \cite{SaTa07},
\begin{equation}
  \label{eq:conslaw}
  f_T = - J_X \, , \quad \text{where} \quad J = - f_X + |A|^2 .
\end{equation}
Taking long-time averages, $\tave{J}_X = - \tave{f_T} = 0$, so in
statistical equilibrium, the time-averaged flux $J$ is uniform in $X$,
$\tave{J} = \tave{-f_X + |A|^2} \equiv \mtwo$.  Integrating over the
domain and using periodicity, we find
\begin{equation}
  \label{eq:m2Amsq}
  \mtwo = \mtwo(L) = \frac{1}{L} \int_0^L \tave{|A(X,\cdot)|^2} \, dX
  > \ 0 . 
\end{equation}
Now for a stationary amplitude death domain, where $\tave{|A|^2} = 0$,
we have $\mtwo = - \tave{f_X}$; this confirms that amplitude death can
occur only where $f$ is \emph{decreasing} on average.  
(In fact, due to the apparent separation of time scales between the
rapid fluctuations $\ffluct$ (on $\bigO(1)$ times) and the slow
overall drift of the mean profile $\fmean(X)$, short-time averaging
seems sufficient to conclude $\shorttave{f_T} \approx 0$, so
that the mean mid-front slope is $\fmean_X(L/2) = -\mtwo$.)
By \eqref{eq:m2Amsq} the (averaged) slope $-\mtwo$ of the large-scale
mode $f$ in the center of the amplitude death state thus seems to be
\emph{globally} determined, being balanced by the mean-square
amplitude of the pattern mode $A$ due to chaotic dynamics concentrated
outside the front region.

Note that also $\mtwo = \tave{|A|^2}$ wherever $\tave{f_X} = 0$,
relating the front slope to the fluctuations in $A$ at extrema of the
averaged profile.  Indeed, in the absence of $A$, $f$ satisfies a heat
equation, and thus by \eqref{eq:conslaw} flows away from local maxima
and towards local minima, destabilizing the front; the added forcing
term in \eqref{eq:MC2} when $|A|^2_X \not= 0$ increases the flux $J$
so as to maintain the averaged overall flux balance, thereby
stabilizing the (averaged) local extrema.  The stabilization mechanism
in the MC equations thus appears to act globally \footnote{Our
  findings are inconsistent with the local stability criterion
  (neglecting the sign of $f_X$) proposed in \cite{SaTa07}, that the
  amplitude death state is stable when the gradient of $f$ is
  sufficiently large, $|f_X| > f_{0c}$ for $f_{0c} \approx 0.44$; for
  large $L$ we find $\mtwo \lesssim \bigO(1/L)$ and observe stable
  fronts, for instance, with $\mtwo \lesssim 0.25$ for $L \gtrsim
  1000$ (see Fig.~\ref{fig:m1m2slopes}(b)).}, with the chaotic
dynamics being essential to sustaining the front (and amplitude death
state); we denote the observed structures \emph{``chaos-stabilized
  fronts''}.

(We remark that \eqref{eq:MC1} does not contain the usual stabilizing
GL cubic term, permitting \eqref{eq:MC1}--\eqref{eq:MC2} to support a
family of exponentially growing solutions $A(X,T) = A_0 e^{T}$,
$f(X,T) = 0$; that is, the MC equations do not have a bounded global
attractor.  However, these growing solutions are dynamically unstable,
in the sense that they are overtaken by faster-growing spatially
varying perturbations \cite{MaCo00}; and in our numerical simulations
we have not observed such  solutions.)


To investigate the behavior of the MC equations on spatially periodic
domains systematically, we have numerically integrated
\eqref{eq:MC1}--\eqref{eq:MC2} using a pseudospectral method in space
and an exponential time differencing (ETDRK4) scheme with step size $H
= 0.02$.  The domain length $L$, the only free parameter in the
system, was chosen to be $L = 2\pi \times 64m/10$ for integers $m$
ranging from 2 to 64; correspondingly, we used between $2^9$ and
$2^{14}$ Fourier modes.  In computing time averages, we integrated
until the system reached a statistically stationary single-front
state, and then averaged over $10^3$-$10^4$ snapshots separated
typically by time intervals $\Delta T = 10$.  All averaging was done
within the frame of reference of the front; that is, we first
determined the front displacement $\sfront(T)$ and used it to align
$A$ and $f$ so that the front was centered at $X = L/2$.  In
particular, the mean profiles were computed by $\fmean(X) =
\tave{f(X+\sfront(\cdot),\cdot)}$.


\paragraph{Averaged profiles:}

As seen in Fig.~\ref{fig:fmeanL}, the averages of $f$ and $|A|$
are, respectively, odd and even about $X = L/2$, recovering the
reflection symmetry of the underlying PDEs
\eqref{eq:MC1}--\eqref{eq:MC2}.  More interestingly, though, the
time-averaged profiles $\fmean(X)$ for $L \gtrsim \Li$
depend strongly on $L$, with the behavior falling into
three distinct regimes:
\begin{figure}
  \begin{center}
    \begin{tabular}{c}
      \includegraphics[width=3.1in]{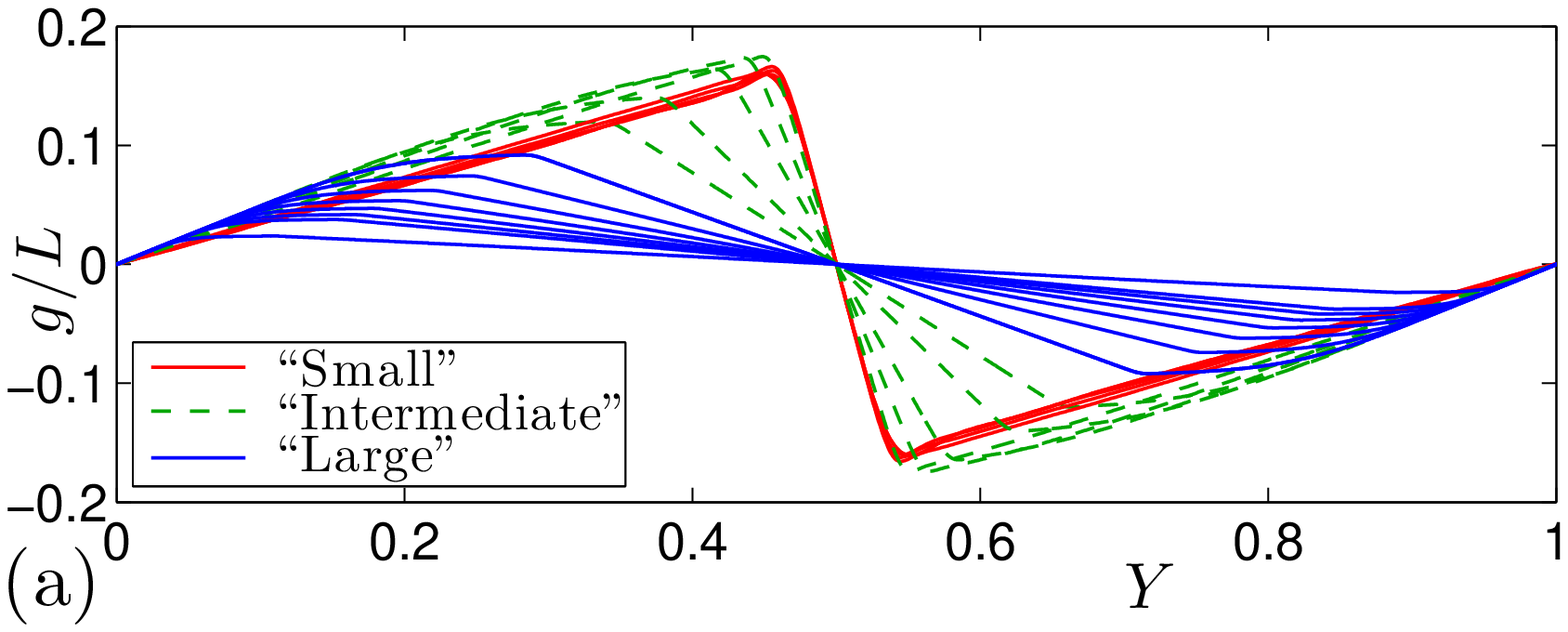}
      \\
      \includegraphics[width=3.1in]{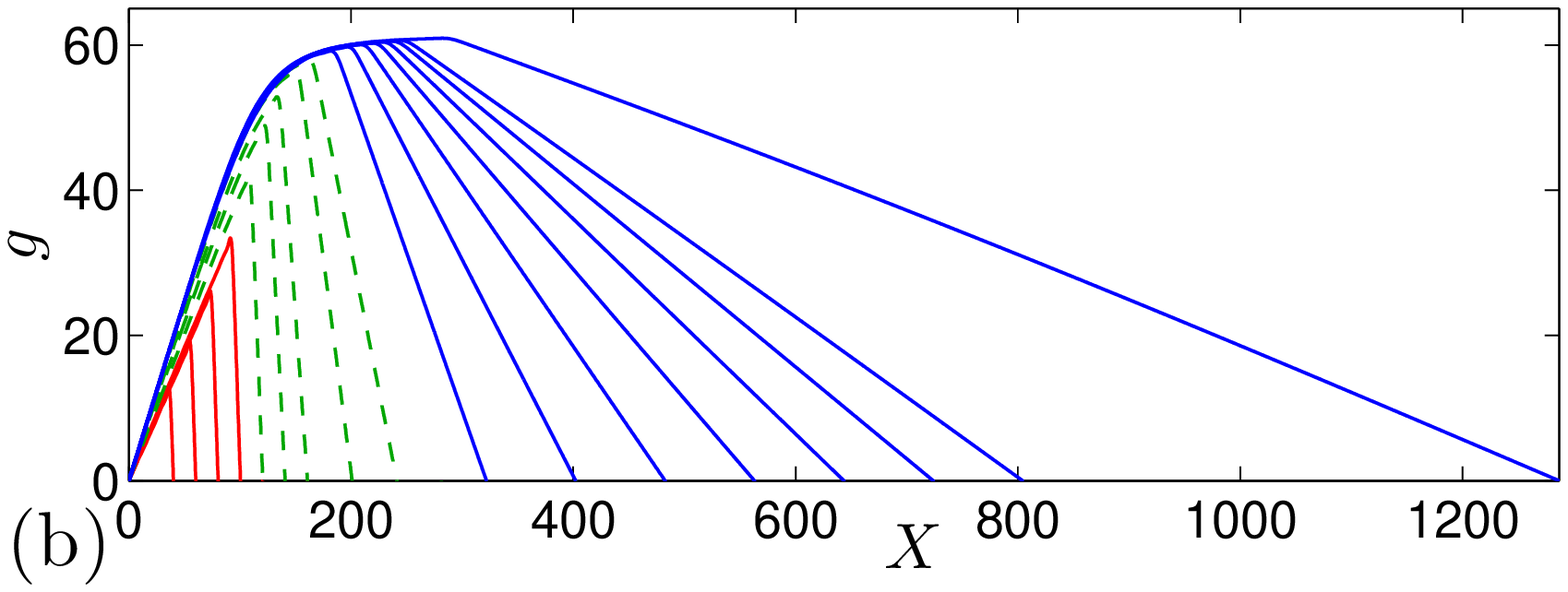} 
      \\
      \includegraphics[width=3.1in]{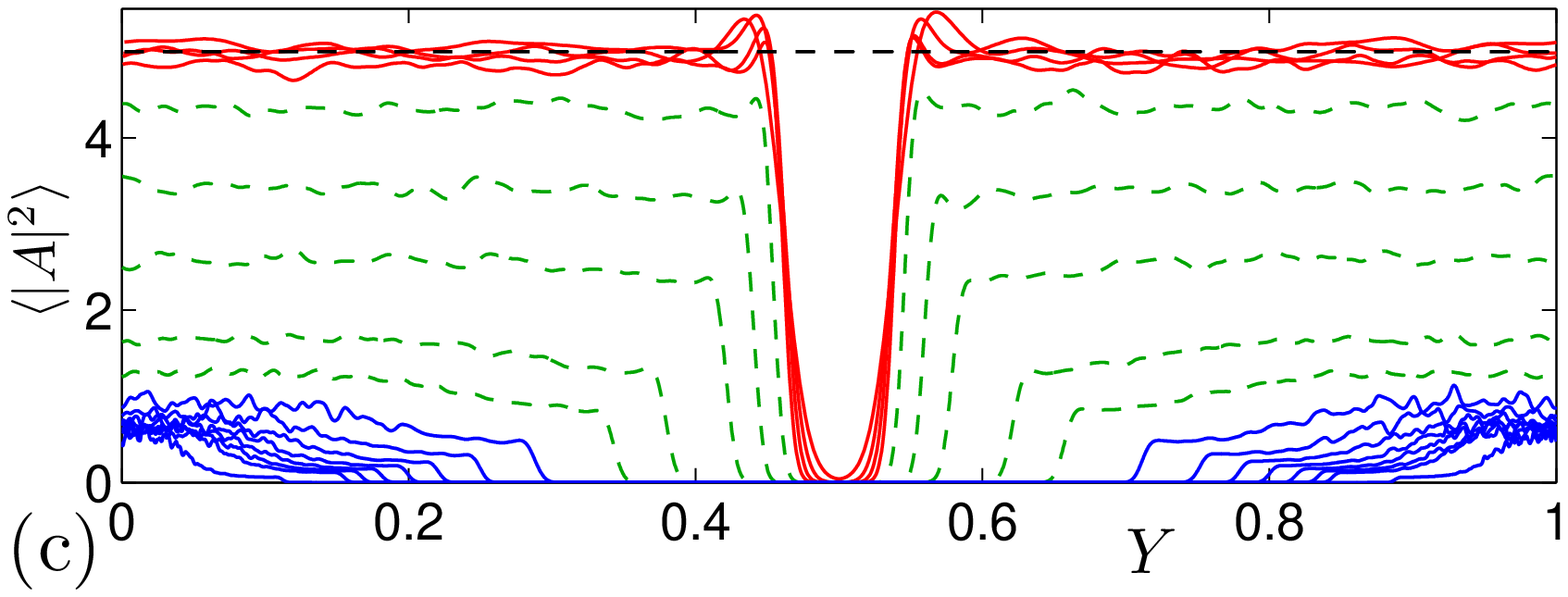}
   \end{tabular}
    \caption{(a), (b) Two representations of the
      long-time-averaged (centered) profile $\fmean(X) =
      \tave{f(X+\sfront(\cdot),\cdot)}$ of the large-scale mode $f$
      for various $L$: (a) scaled horizontally and vertically, $L^{-1}
      \fmean(L \Xscal)$ for $\Xscal = X/L \in [0,1]$; (b) unscaled,
      half of the (odd) profile, $\fmean(X)$ for $X \in [0,L/2]$.  (c)
      Centered time-averaged pattern amplitude $\tave{|A(L \Xscal,
        \cdot)|^2}$.  Domain sizes are (``small'': red)
      $L = {25.6\pi}$, ${38.4\pi}$, ${51.2\pi}$, ${64\pi}$;
      (``intermediate'': green, dashed lines) $L = {76.8\pi}$, ${89.6\pi}$,
      ${102.4\pi}$, ${128\pi}$, ${153.6\pi}$; and (``large'': blue) $L
      = {204.8\pi}$, ${256\pi}$, ${307.2\pi}$, ${358.4\pi}$,
      ${409.6\pi}$, ${460.8\pi}$, ${512\pi}$ and ${819.2\pi}$.
      Averages were taken over time periods $T = 1\times 10^5$ (small,
      intermediate) and $2\times 10^4$ (large), with $\Delta T = 10$
      between snapshots.\\[-6ex]}
  \label{fig:fmeanL}
  \end{center}
\end{figure}

\begin{figure}
  \begin{center}
    \begin{tabular}{c}
      \includegraphics[width=3.2in]{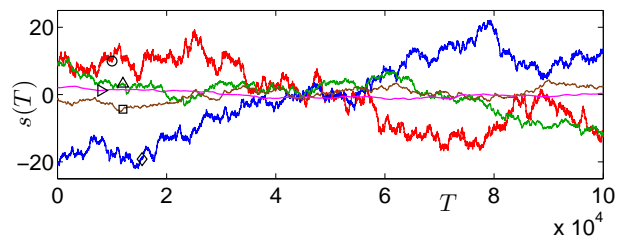}
    \end{tabular}
    \caption{Representative trajectories of the front displacement
      $\sfront(T)$ about $X=L/2$ for the ``small'' domains $L =
      38.4\pi\;(\circ)$ and $51.2\pi\;(\Diamond)$, and
      ``intermediate'' sizes $L = 76.8\pi\;(\vartriangle)$,
      ${102.4\pi}\;(\Box)$, and
      ${153.6\pi}\;(\vartriangleright)$.\\[-6ex]}
    \label{fig:frontpos}
  \end{center}
\end{figure}

For relatively \emph{``small'' domains}, $\Li \lesssim L \lesssim \Lii
\approx 220$, the scaled profiles in Fig.~\ref{fig:fmeanL}(a)
approximately coincide, indicating a scaling form for $\fmean$: for
some fundamental shape function $\fscal$, periodic on $[0,1]$, we have
$\fmean(X) \approx L \fscal(\Xscal)$ (with $\Xscal = X/L$).  
In this ``small-$L$'' regime the scaling relation is highly accurate
within the front (but is weakly violated outside it: the slope $\mone
= \fmean_X(0)$ in the active region increases slowly with $L$; see
Fig.~\ref{fig:m1m2slopes}(a)); in particular, the midpoint slope is
independent of $L$, with $-\mtwo = \fmean_X(L/2) = \fscal'(0.5)
\approx -4.6$ (cf.\ \cite{SaTa07}); see Fig.~\ref{fig:m1m2slopes}(b), where
we have also numerically verified \eqref{eq:m2Amsq}.
\begin{figure}
  \begin{center}
    \begin{tabular}{cc}
      \includegraphics[width=0.23\textwidth]{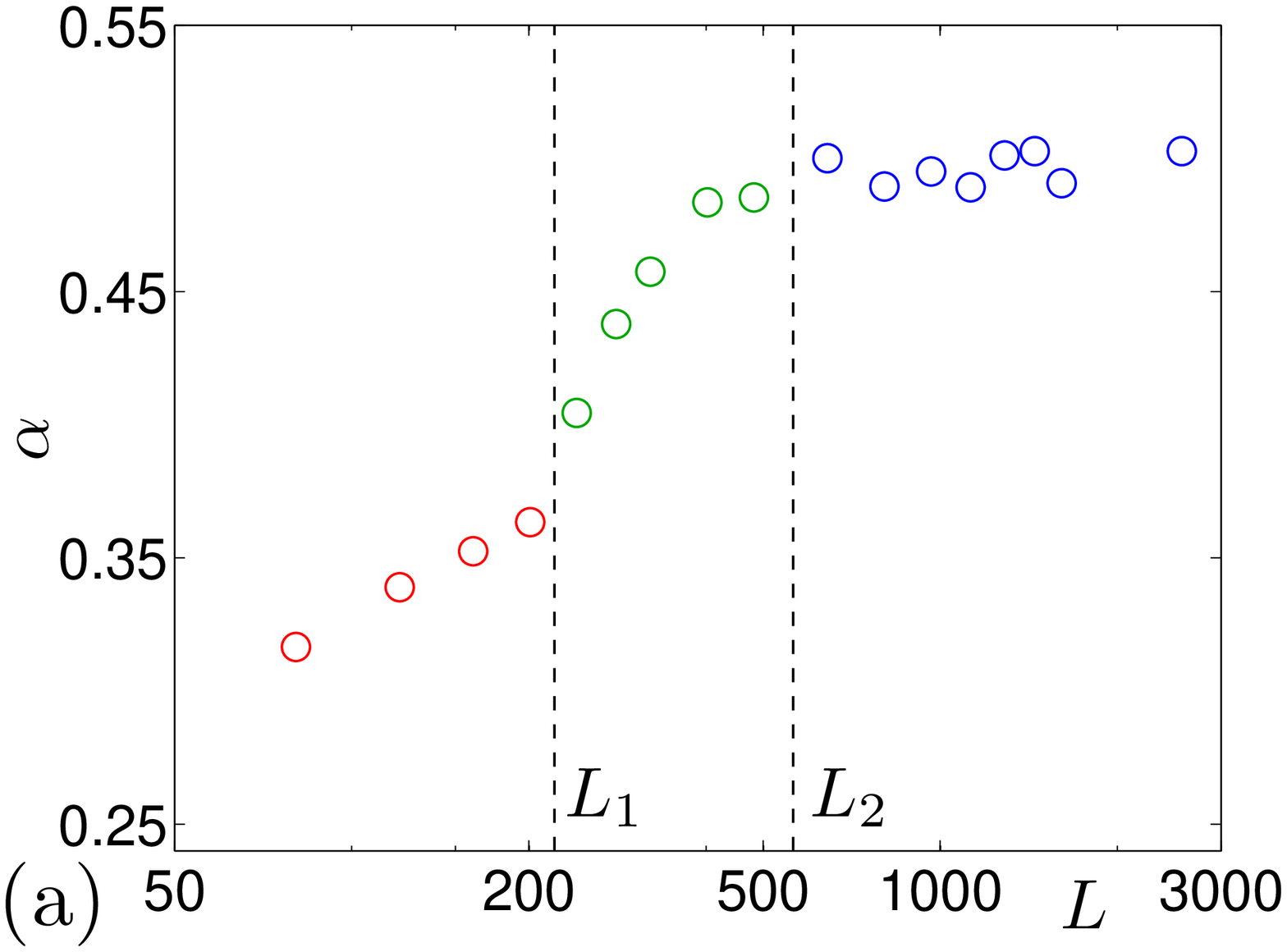}
      \includegraphics[width=0.23\textwidth]{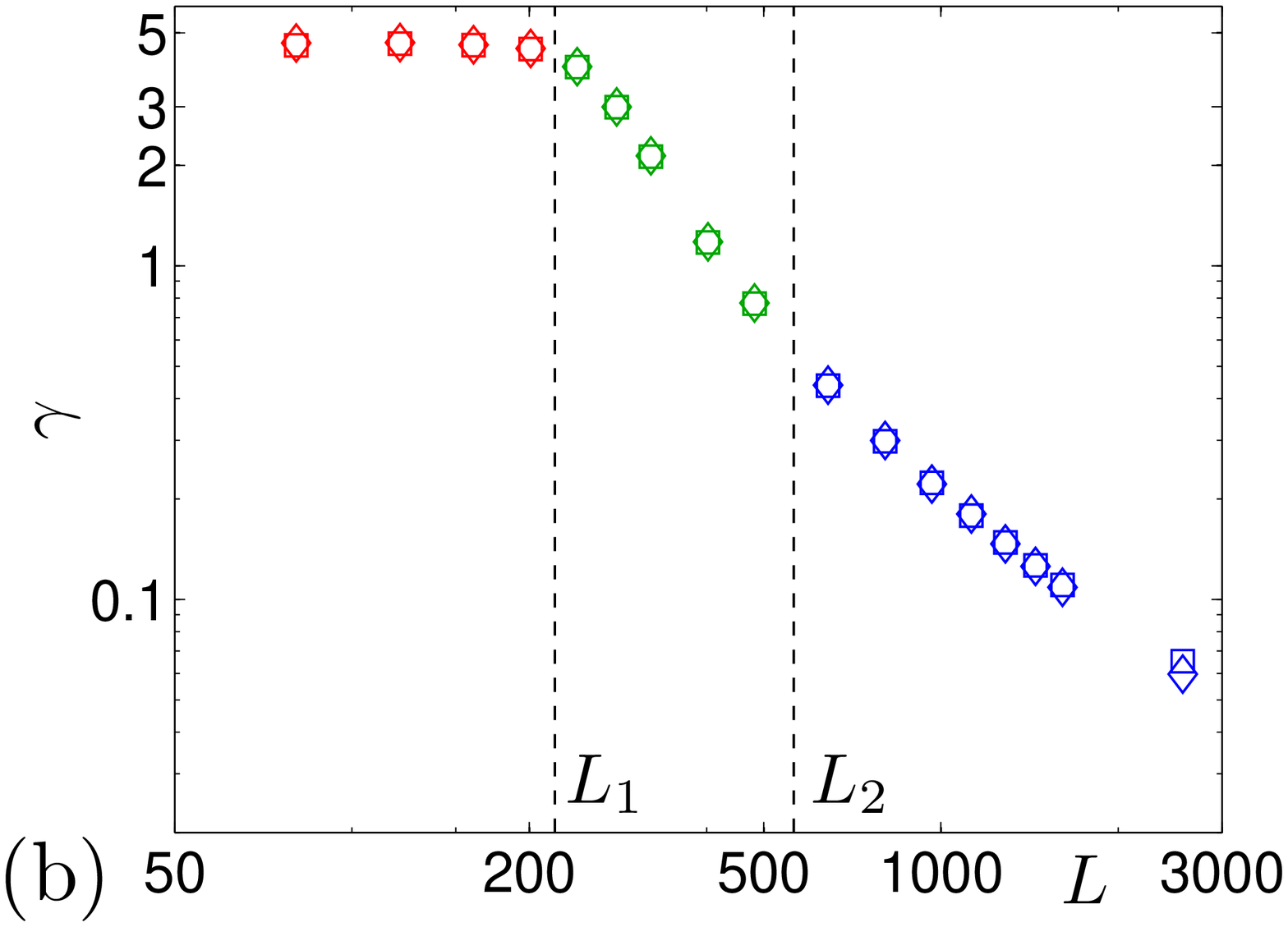}
    \end{tabular}
    \caption{(a) Slope $\mone = \fmean_X(0)$ of averaged profile at
      midpoint of chaotic region.  (b) Absolute value of mid-front
      slope $\mtwo = - \fmean_X(L/2)$ $(\Box)$, shown with the
      mean-square average pattern amplitude $L^{-1} \int_0^L
      \tave{|A(X,\cdot)|^2} \, dX$ $(\Diamond)$, verifying
      \eqref{eq:m2Amsq}.  Lengths $L$ and colors are as in
      Fig.~\ref{fig:fmeanL}; the vertical lines at $\Lii \approx 220$
      and $\Liii \approx 560$ indicate approximate transitions between
      ``small'', ``intermediate'' and ``large'' regimes.\\[-6ex]}
  \label{fig:m1m2slopes}
  \end{center}
\end{figure}%
We also find that the relative sizes of the front and chaotic regions
remain fixed, with $\tave{|A|^2}(X) \approx 5$ approximately constant
and $L$-independent in the chaotic region (Fig.~\ref{fig:fmeanL}(c)).
For these ``small'' domain sizes, the front translates over long times
(recall Fig.~\ref{fig:fAxt}); interestingly, the statistics of the
front motion appear consistent with a random walk \cite{Poon09p}, as
suggested by the trajectories of $\sfront(T)$ shown in
Fig.~\ref{fig:frontpos}.%

The (approximate) scaling form for the time-averaged profile
$\fmean(X)$ observed for ``small'' domains breaks down for larger $L$.
Instead, for domain sizes in an \emph{``intermediate'' regime} with
lengths $\Lii \lesssim L \lesssim \Liii \approx 560$, the amplitude of
$\fmean$ begins to level off, the front becomes wider and less steep,
and chaotic fluctuations of $A$ and $f$ decrease in amplitude (see
Figs.~\ref{fig:fmeanL}--\ref{fig:m1m2slopes}).  Furthermore, the
variance of the front displacement $\sfront(T)$ decreases strongly
with $L$, until the translation becomes imperceptible
(Fig.~\ref{fig:frontpos}).

This behavior is transitional to that of \emph{``large'' domains} $L
\gtrsim \Liii \approx 560$.  In this regime the front is stationary,
$\sfront(T) \equiv 0$; the amplitude of $\fmean(X)$ saturates at $\max
\fmean \approx 62$, as does the maximum slope in the chaotic region,
$\mone = \fmean_X(0) \approx 0.5$.  Indeed, Fig.~\ref{fig:fmeanL}(b)
shows that the mean profile $\fmean(X)$ near $X = 0$ becomes invariant
with increasing $L$; this saturation of the profile indicates to us
that we have reached the large-$L$ asymptotic regime of
\eqref{eq:MC1}--\eqref{eq:MC2}.  Since the width of the amplitude
death region continues to grow with $L$, while the height is bounded,
the front slope $-\mtwo$ decays with $L$, and hence so does the
amplitude of the fluctuations in $A$: for large $L$ the spatially
localized chaotic dynamics superimposed on the mean profile are
strongly suppressed.


\paragraph{Transient behavior:}

\begin{figure}
  \begin{center}
    \begin{tabular}{c}
      \includegraphics[width=3.2in]{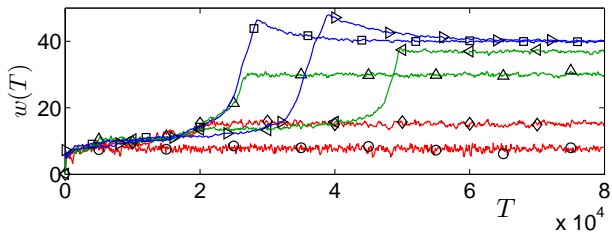}
      \\ \mbox{} \vspace{-6ex}
      \end{tabular}
  \end{center}
  \caption{Evolution of $w(T) = [ L^{-1} \int_0^L f(X,T)^2 \, dX
    ]^{1/2}$ for the ``small'' domains $L = {25.6\pi\;(\circ)}$ and
    ${51.2\pi\;(\Diamond)}$, ``intermediate'' domains $L =
    {89.6\pi\;(\vartriangle)}$ and ${128\pi\;(\vartriangleleft)}$; and
    ``large'' domains $L = {256\pi\;(\Box)}$ and
    ${307.2\pi\;(\vartriangleright)}$, computed to $T = 8
    \times 10^4$.\\[-2ex]}
  \label{fig:width}
\end{figure}

\begin{figure}
  \begin{center}
    \begin{tabular}{cc}
      \includegraphics[width=0.23\textwidth]{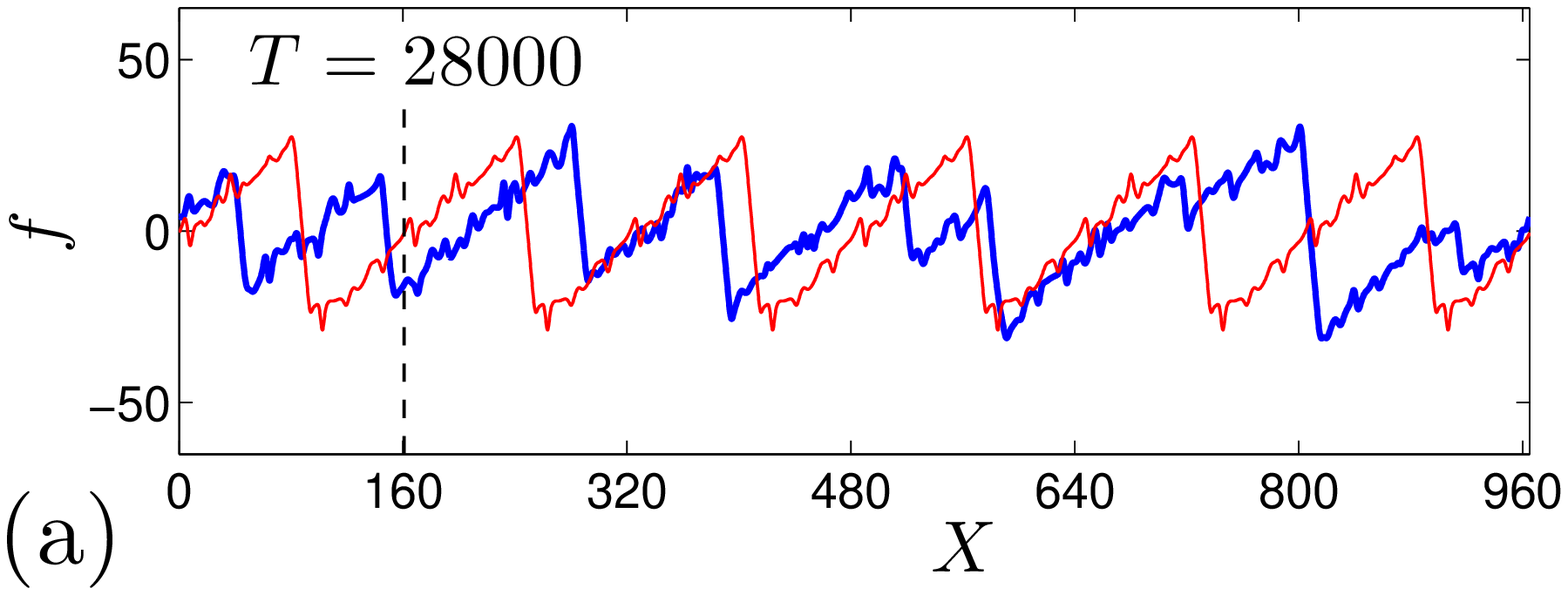}
      & 
      \includegraphics[width=0.23\textwidth]{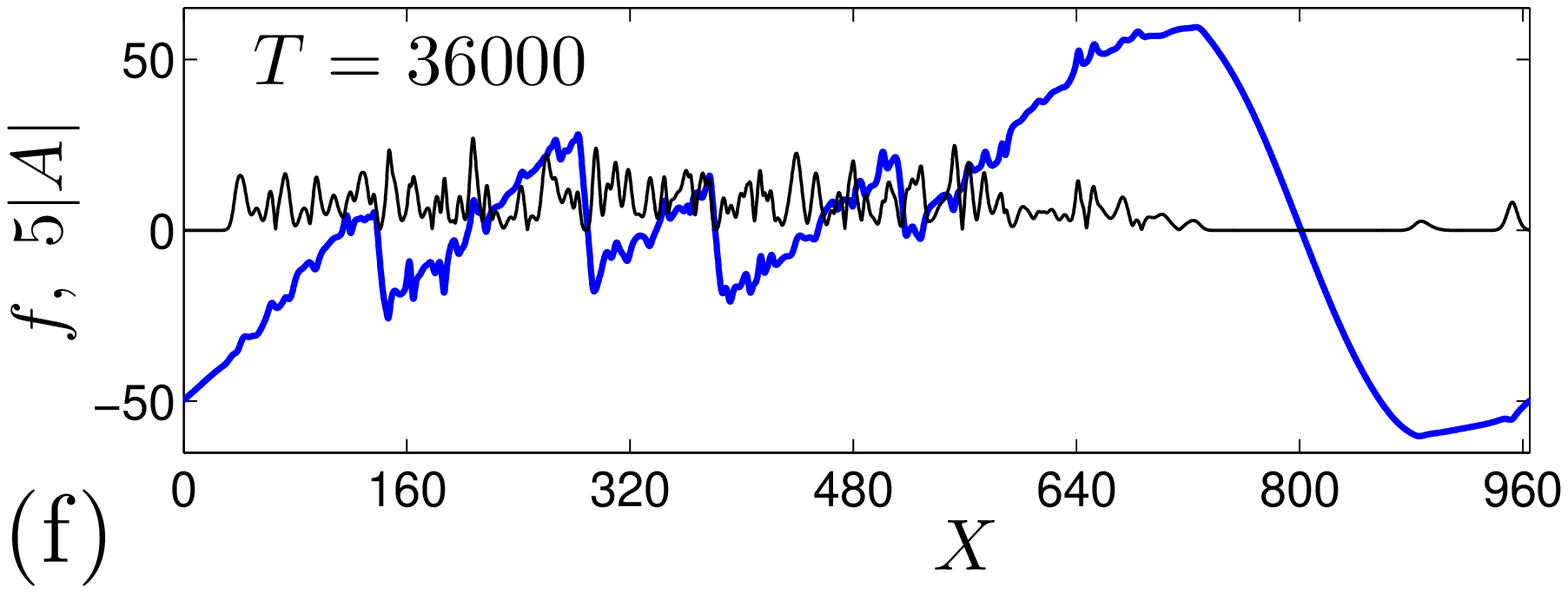}
      \\
      \includegraphics[width=0.23\textwidth]{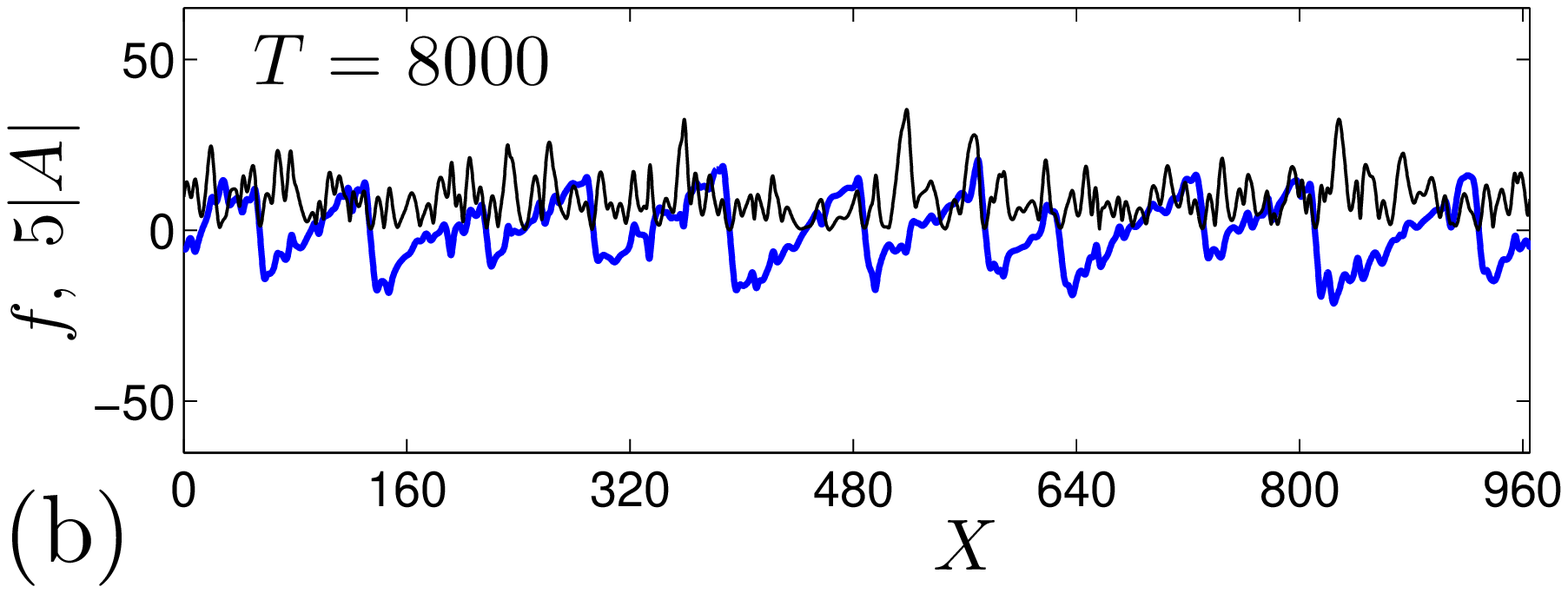}
      &
      \includegraphics[width=0.23\textwidth]{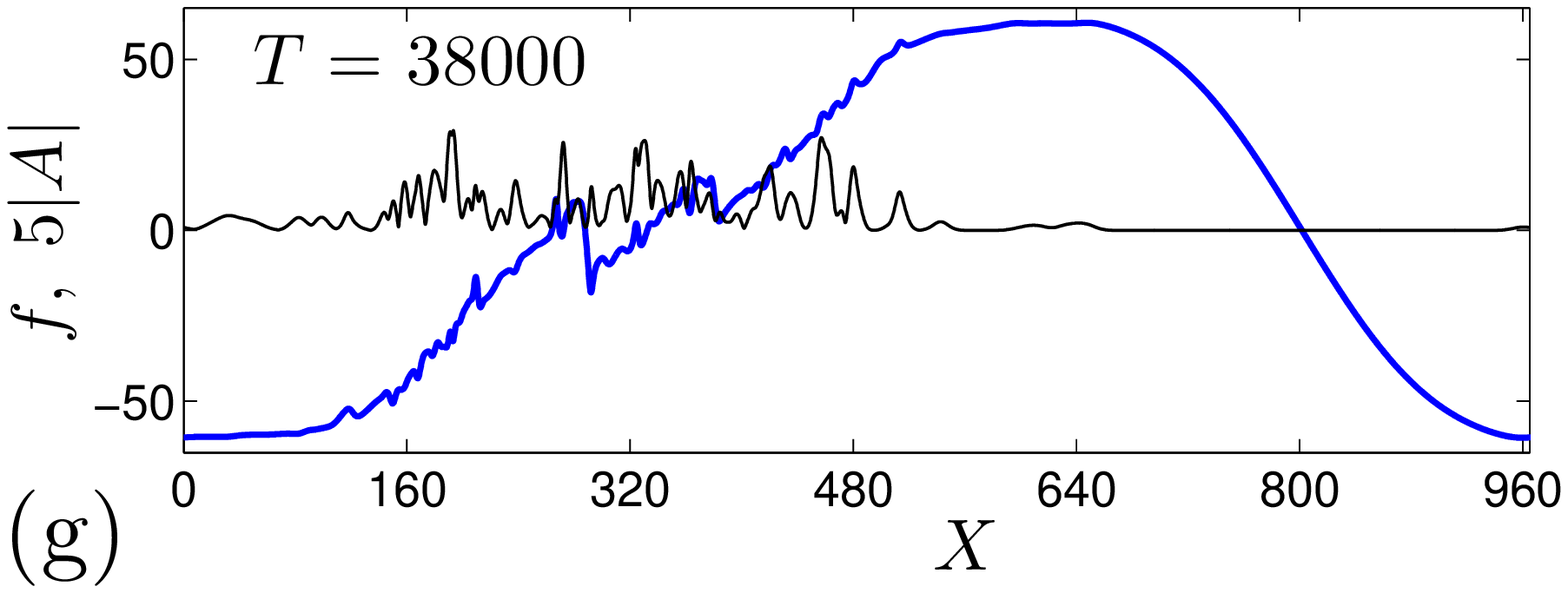}
      \\
      \includegraphics[width=0.23\textwidth]{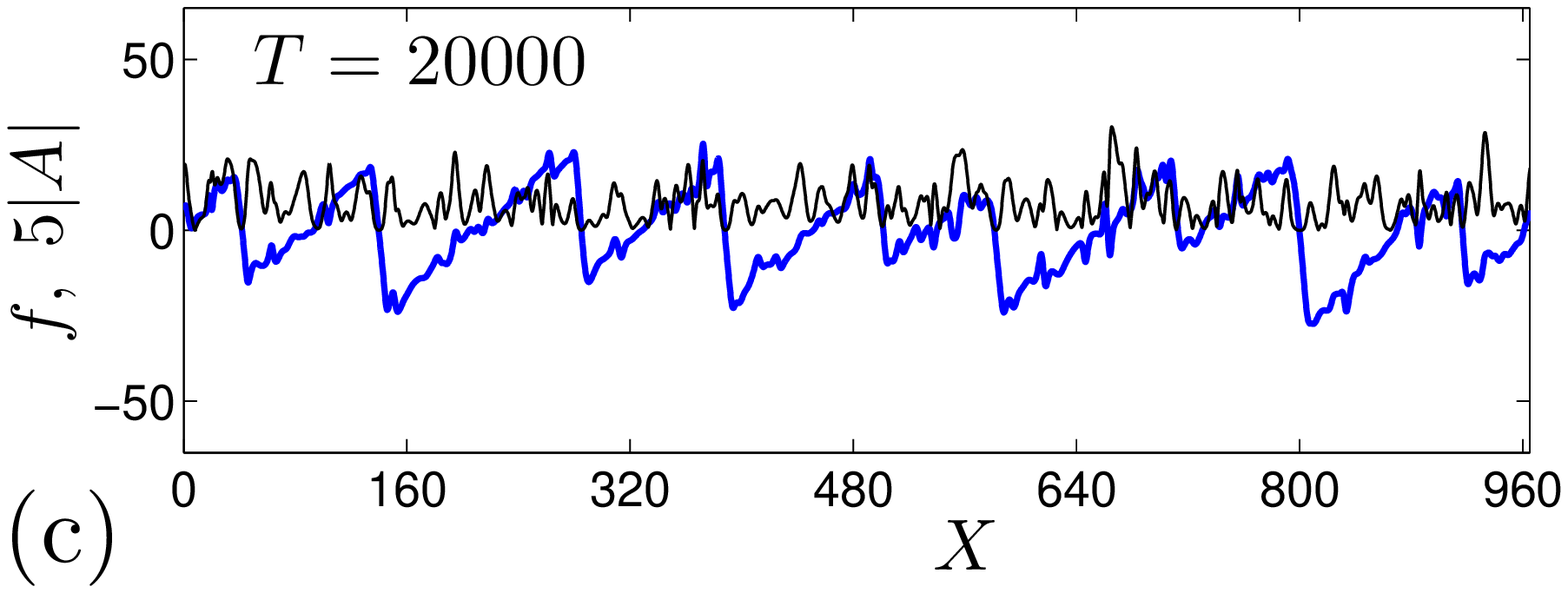}
      &
      \includegraphics[width=0.23\textwidth]{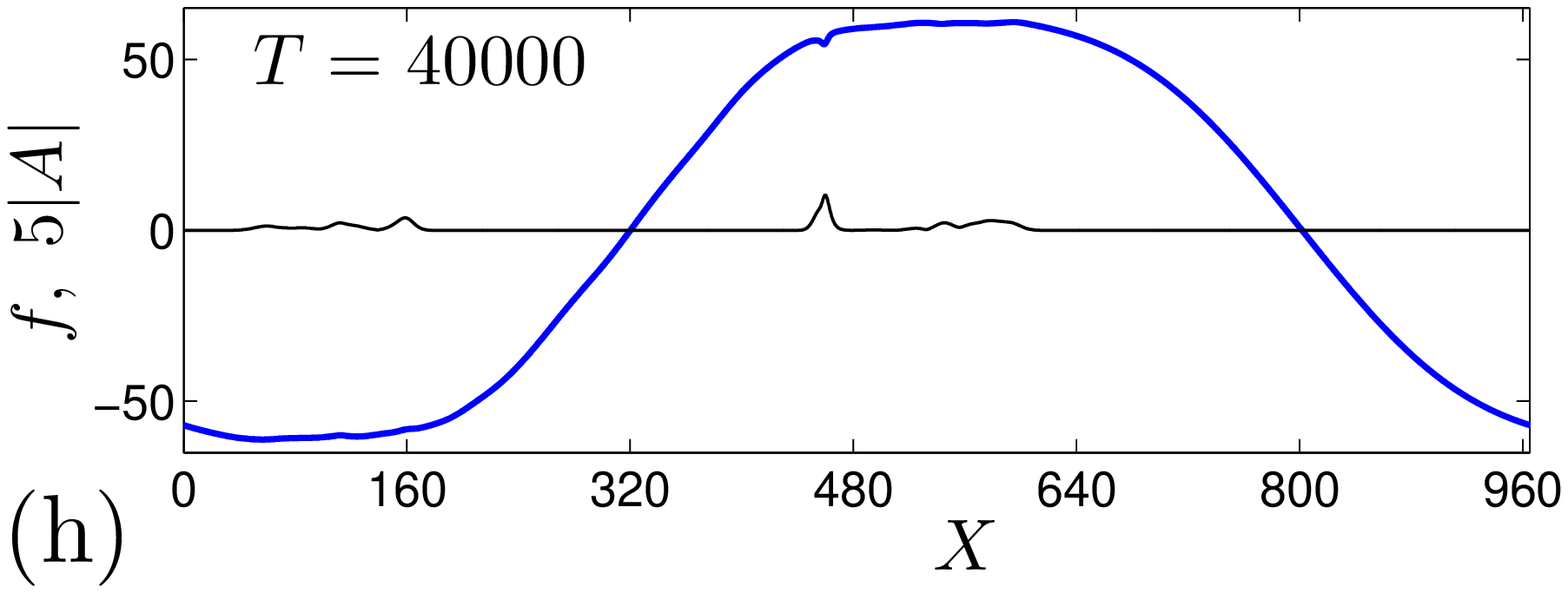}
      \\
      \includegraphics[width=0.23\textwidth]{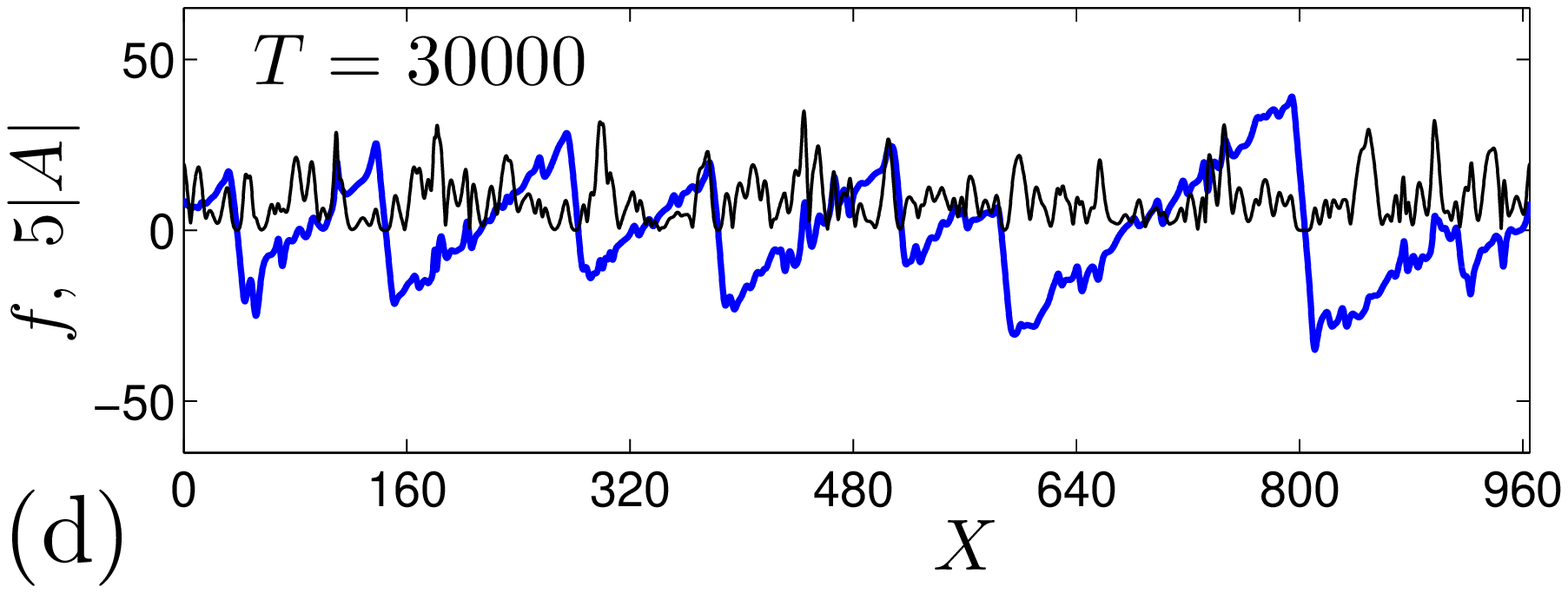}
      &
      \includegraphics[width=0.23\textwidth]{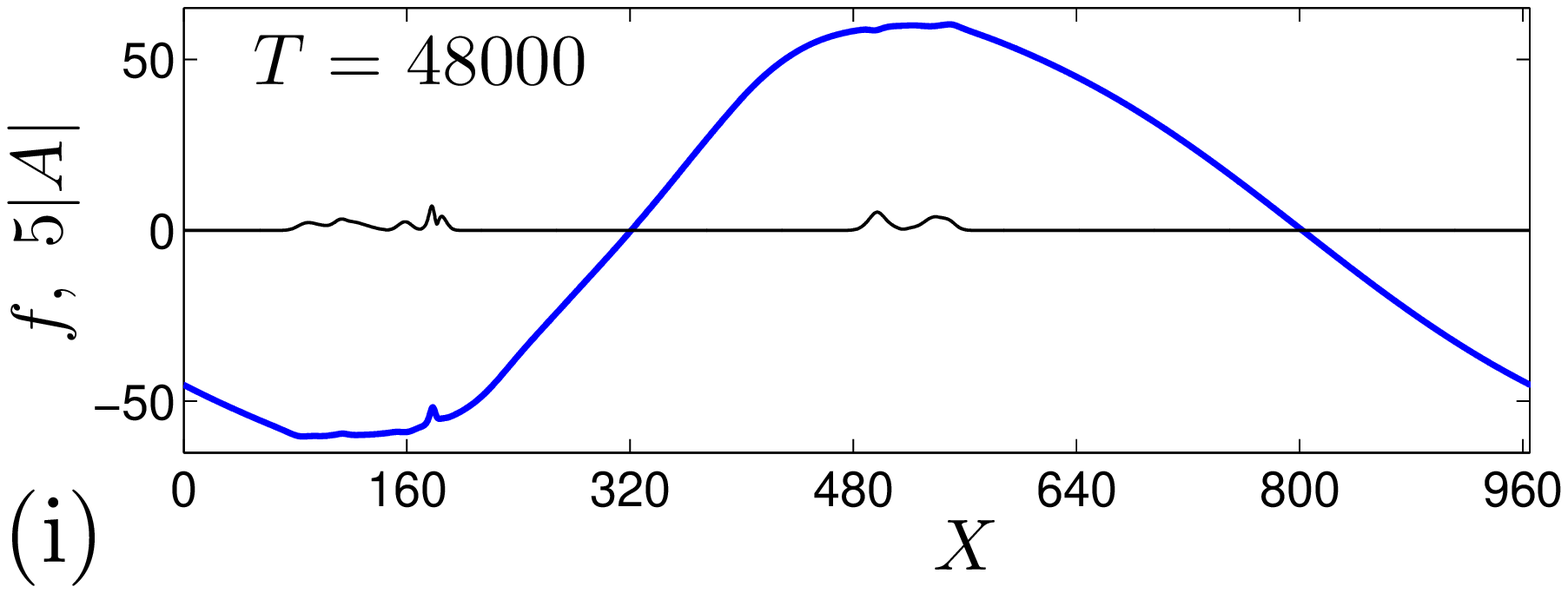}
      \\
      \includegraphics[width=0.23\textwidth]{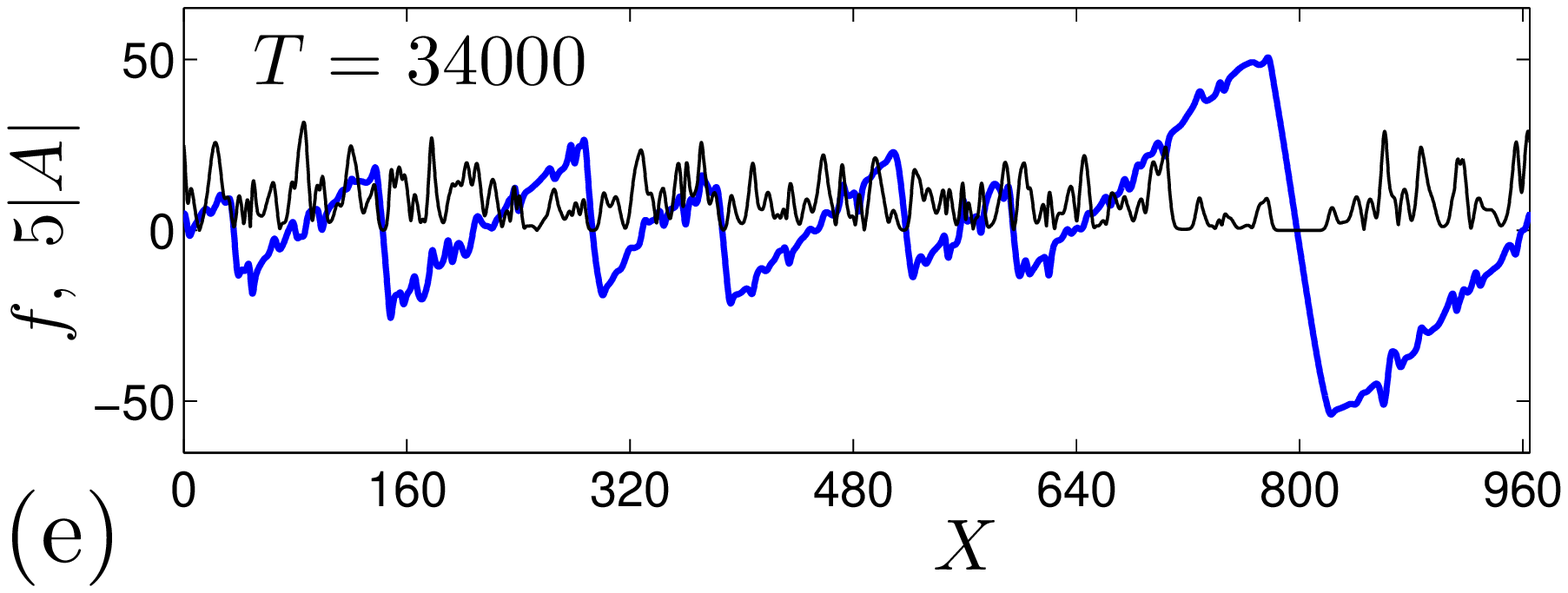}
      &
      \includegraphics[width=0.23\textwidth]{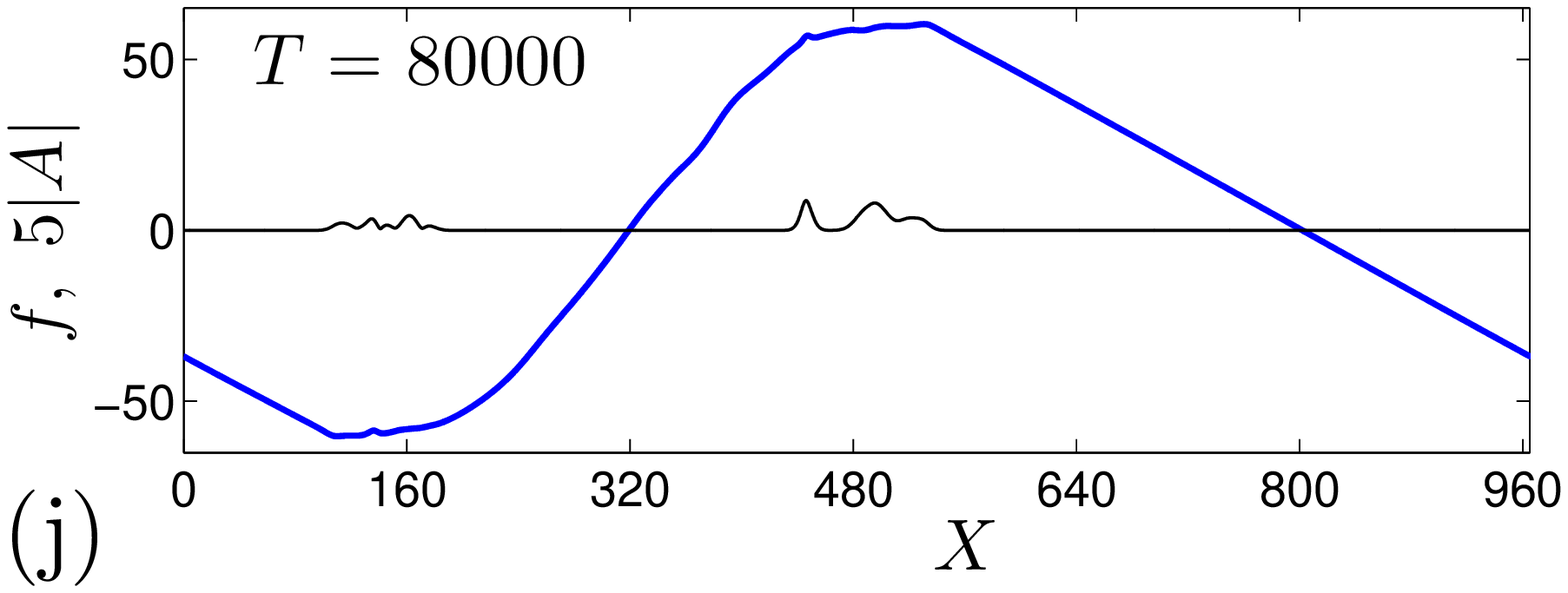}
      \mbox{}
    \end{tabular}
    \caption{(a) Large scale mode $f$ at $T = 28000$ for the ``large''
      domain $L = 307.2\pi$ with, for comparison, six copies of the $L
      = 51.2\pi$ profile from Fig.~\ref{fig:snapshot}.  (b)--(j)
      Snapshots of $f$ and $|A|$ (for clarity we plot $5|A|$) showing
      coarsening and collapse to a single front for $L = 307.2\pi
      \approx 964.8$.\\[-7ex]}
    \label{fig:coarseningts}
  \end{center}
\end{figure}

The strong $L$-dependence of the properties of the MC equations,
within identifiable domain size regimes, is apparent also in the
transient approach to the long-time statistically stationary state, as
summarized in the time evolution of $w(T) = [ L^{-1} \int_0^L f(X,T)^2
\, dX ]^{1/2}$ --- analogous to an interface width in the context of
surface growth --- as in Fig.~\ref{fig:width}.
The snapshots from a typical time evolution for a ``large'' domain in Fig.~\ref{fig:coarseningts} 
demonstrate an extended coarsening period followed by a remarkable
collapse to a single front:

From small random data, initial growth rapidly establishes a sawtooth
pattern in $f$: a concatenation of structures, of varying widths and
corresponding heights, locally reminiscent of the statistically
stationary states in ``small'' domains (see
Fig.~\ref{fig:coarseningts}(a)).  Once this metastable state of
multiple Burgers-like viscous shocks with superimposed chaotic
fluctuations is established, a slow coarsening process ensues: front
structures grow and merge with adjacent fronts, leading to a gradual
increase of length scales and of $w(T)$
(Fig.~\ref{fig:coarseningts}(b)--(d)).

For ``small'' domains, for which the long-time state has the
(approximate) scaling form $\fscal$ on average, this coarsening
concludes once there is a single front.  However, for ``intermediate''
and ``large'' domains, the gradual growth of $w(T)$ through coarsening
is followed by a ``jump'' in $w(T)$ (see Fig.~\ref{fig:width})
\cite{Poon09p}, reflecting qualitative changes in the profile
$f(X,T)$, as seen in Fig.~\ref{fig:coarseningts}(d)--(h).
Specifically, having (presumably) exceeded a critical size, one of the
front structures begins to dominate, and then grows relatively rapidly
by engulfing its neighbors until a state with an $L$-dependent single
front is attained.

Finally, for ``large'' domains, $w(T)$ overshoots its asymptotic value
(Fig.~\ref{fig:width}), since following the collapse to a single
front, $f$ is initially nonlinear in the amplitude death region,
before undergoing slow diffusive relaxation (by \eqref{eq:MC2} with
$|A| = 0$) to the time-asymptotic linear front profile
(Fig.~\ref{fig:coarseningts}(h)--(j)).


\paragraph{Discussion:}

The structure of the Matthews-Cox equations
\eqref{eq:MC1}--\eqref{eq:MC2} is reminiscent of that of other
well-known systems.  For instance, viewing \eqref{eq:MC2} as a heat
equation for $f$ with (localized chaotic) forcing, using the heat
kernel to express $f$ as a quadratic functional of $A$ and
substituting, the $-\ii f A$ coupling term in \eqref{eq:MC1} acts as a
nonlocal cubic stabilizing term in a GL-type equation.  Alternatively,
in the light of the viscous shock-like behavior in $f$, it may be
fruitful to view \eqref{eq:MC2} as a generalized viscous Burgers
equation, with a nonlocal forcing term determined by \eqref{eq:MC1}.
Such considerations may facilitate a theoretical understanding of the
unusual behavior we have described in the MC equations.


\vspace{2ex}


  We thank the IRMACS Centre at Simon Fraser University for providing
  a productive research environment, and David Muraki and Richard
  Koll\'ar for helpful discussions.  This work was partially supported by NSERC.


\newcommand{\SortNoop}[1]{}

\end{document}